\documentclass[10pt,a4paper]{article}
\usepackage{amsmath}
\usepackage{amssymb}
\usepackage{amsfonts}
\usepackage{amstext}
\usepackage{amsbsy}
\usepackage[mathscr]{eucal}
\usepackage{graphicx}
\usepackage{color}
\newcommand{\beq}{\begin{equation}}
\newcommand{\eeq}{\end{equation}}
\newcommand{\beqa}{\begin{eqnarray}}
\newcommand{\eeqa}{\end{eqnarray}}
\newcommand{\ket}[1]{| #1 \rangle}

\makeindex
\title{\Large\textbf{General pure multipartite entangled states and the Segre variety}}

\author{\textit{ Hoshang Heydari}\\
        \small\textit{Institute of Quantum
Science, Nihon University,}\\
\small\textit{1-8 Kanda-Surugadai, Chiyoda-ku, Tokyo 101-8308, Japan
}}

\date{}
\begin{document}

\maketitle \thispagestyle{empty}

\begin{abstract}
In this paper, we construct a measure of entanglement by
generalizing the quadric polynomial of the Segre variety for general
multipartite states. We give explicit expressions for general pure
three-partite and four-partite states. Moreover, we will discuss and
compare this measure of entanglement with the generalized
concurrence.
\end{abstract}

\section{Introduction}
Quantum entanglement has received a lot of attention during the
recent years because of its usefulness in many quantum information
and communication tasks such as quantum teleportation and quantum
cryptography. However, there is still  many open questions
concerning the quantification and classification  of multipartite
states and also its true nature. Thus a deep understanding of this
interesting quantum mechanical phenomena  could result in
construction of new algorithms and protocols for quantum information
processing.  One widely used measure of entanglement is the
so-called concurrence \cite{Wootters98}.  The connection between
concurrence and geometry is found in a map called the Segre
embedding \cite{Dorje99,Miyake}. In this paper we will expand our
result on the Segre variety \cite{Hosh5}, which is a quadric space
in algebraic geometry, by constructing a measure of entanglement for
general pure multipartite states, which also coincides with
concurrence of a general pure bipartite and three-partite states
\cite{Albeverio}.
 Now, let us start by denoting
a general, pure, composite quantum system with $m$ subsystems as
$\mathcal{Q}=\mathcal{Q}^{p}_{m}(N_{1},N_{2},\ldots,N_{m})
=\mathcal{Q}_{1}\mathcal{Q}_{2}\cdots\mathcal{Q}_{m}$, consisting of
the pure state $
\ket{\Psi}=\sum^{N_{1}}_{k_{1}=1}\sum^{N_{2}}_{k_{2}=1}\cdots\sum^{N_{m}}_{k_{m}=1}
$ $\alpha_{k_{1},k_{2},\ldots,i_{m}} \ket{k_{1},k_{2},\ldots,k_{m}}
$ and corresponding to the Hilbert space $
\mathcal{H}_{\mathcal{Q}}=\mathcal{H}_{\mathcal{Q}_{1}}\otimes
\mathcal{H}_{\mathcal{Q}_{2}}\otimes\cdots\otimes\mathcal{H}_{\mathcal{Q}_{m}}
$, where the dimension of the $j$th Hilbert space is
$N_{j}=\dim(\mathcal{H}_{\mathcal{Q}_{j}})$. We are going to use
this notation throughout this paper. In particular, we denote a pure
two-qubit state by $\mathcal{Q}^{p}_{2}(2,2)$. Next, let
$\rho_{\mathcal{Q}}$ denote a density operator acting on
$\mathcal{H}_{\mathcal{Q}}$. The density operator
$\rho_{\mathcal{Q}}$ is said to be fully separable, which we will
denote by $\rho^{sep}_{\mathcal{Q}}$, with respect to the Hilbert
space decomposition, if it can  be written as $
\rho^{sep}_{\mathcal{Q}}=\sum^\mathrm{N}_{k=1}p_k
\bigotimes^m_{j=1}\rho^k_{\mathcal{Q}_{j}},~\sum^\mathrm{N}_{k=1}p_{k}=1
$
 for some positive integer $\mathrm{N}$, where $p_{k}$ are positive real
numbers and $\rho^k_{\mathcal{Q}_{j}}$ denotes a density operator on
Hilbert space $\mathcal{H}_{\mathcal{Q}_{j}}$. If
$\rho^{p}_{\mathcal{Q}}$ represents a pure state, then the quantum
system is fully separable if $\rho^{p}_{\mathcal{Q}}$ can be written
as
$\rho^{sep}_{\mathcal{Q}}=\bigotimes^m_{j=1}\rho_{\mathcal{Q}_{j}}$,
where $\rho_{\mathcal{Q}_{j}}$ is the density operator on
$\mathcal{H}_{\mathcal{Q}_{j}}$. If a state is not separable, then
it is said to be an entangled state.  For those readers who are
unfamiliar with algebraic geometry,
 we give a
short introduction to the basic definition of complex algebraic and
projective variety. However, the standard references for the complex
projective variety are \cite{Griff78,Mum76}.
Let $\{g_{1},g_{2},\ldots,g_{q}\}$ be continuous functions
$\mathbf{C}^{n}\longrightarrow\mathbf{C}$. Then we define a complex
space as the set of simultaneous zeroes of the functions
\begin{eqnarray}
&&\mathcal{V}_{\mathbf{C}}(g_{1},g_{2},\ldots,g_{q})=\{(z_{1},z_{2},
\ldots,z_{n})\in\mathbf{C}^{n}:\\\nonumber&&
g_{i}(z_{1},z_{2},\ldots,z_{n})=0~\forall~1\leq i\leq q\}.
\end{eqnarray}
The complex space becomes a topological space by giving them the
induced topology from $\mathbf{C}^{n}$. Now, if all $g_{i}$ are
polynomial functions in coordinate functions, then the real
(complex) space is called a real (complex) affine variety. A complex
projective space $\mathbf{CP}^{n}$ is defined to be the set of lines
through the origin in $\mathbf{C}^{n+1}$, that is,
\begin{equation}
\mathbf{CP}^{n}=\frac{\mathbf{C}^{n+1}-{0}}{
(x_{1},\ldots,x_{n+1})\sim(y_{1},\ldots,y_{n+1})},~\lambda\in
\mathbf{C}-0,~x_{i}=y_{i} \forall ~0\leq i\leq n.
\end{equation}
The complex manifold $\mathbf{CP}^{1}$ of dimension one is a very
important one, since as a real manifold it is homeomorphic to the
2-sphere $\mathbf{S}^{2}$,e.g., the Bloch sphere. Moreover, every
complex compact manifold can be embedded in some $\mathbf{CP}^{n}$.
In particular, we can embed a product of two projective spaces into
a third one. Let $\{g_{1},g_{2},\ldots,g_{q}\}$ be a set of
homogeneous polynomials in the coordinates
$\{\alpha_{1},\alpha_{2},\ldots,\alpha_{n+1}\}$ of
$\mathbf{C}^{n+1}$. Then the projective variety is defined to be the
subset
\begin{eqnarray}
&&\mathcal{V}(g_{1},g_{2},\ldots,g_{q})=\{[\alpha_{1},\ldots,\alpha_{n+1}]
\in\mathbf{CP}^{n}:\\\nonumber&&
g_{i}(\alpha_{1},\ldots,\alpha_{n+1})=0~\forall~1\leq i\leq q\}.
\end{eqnarray}
We can view the complex affine variety
$\mathcal{V}_{\mathbf{C}}(g_{1},g_{2},\ldots,g_{q})\subset\mathbf{C}^{n+1}$
as a complex cone over the projective variety
$\mathcal{V}(g_{1},g_{2},\ldots,g_{q})$.

\section{Multi-projective  variety and a multipartite entanglement
measure} In this section, we will review the construction of the
Segre variety. Then, we will construct a measure of entanglement for
general multipartite states based on a modification of the
definition of the Segre variety. This is an extension of our
pervious result on construction of a measure of entanglement for
general pure multipartite states \cite{Hosh5}. We can map the
product of  space $\mathbf{CP}^{N_{1}-1}\times\mathbf{CP}^{N_{2}-1}
\times\cdots\times\mathbf{CP}^{N_{m}-1}$ into a projective variety
by its Segre embedding as follows. Let
$(\alpha_{1},\alpha_{2},\ldots,\alpha_{N_{i}})$  be points defined
on the complex projective space $\mathbf{CP}^{N_{i}-1}$. Then the
Segre map
\begin{equation}
\begin{array}{ccc}
  \mathcal{S}_{N_{1},\ldots,N_{m}}:\mathbf{CP}^{N_{1}-1}\times\mathbf{CP}^{N_{2}-1}
\times\cdots\times\mathbf{CP}^{N_{m}-1}&\longrightarrow&
\mathbf{CP}^{N_{1}N_{2}\cdots N_{m}-1}\\
 ((\alpha_{1},\alpha_{2},\ldots,\alpha_{N_{1}}),\ldots,
 (\alpha_{1},\alpha_{2},\ldots,\alpha_{N_{m}})) & \longmapsto&
 (\ldots,\alpha_{i_{1},i_{2},\ldots, i_{m}},\ldots). \\
\end{array}
\end{equation}
is well defined for $\alpha_{i_{1}i_{2}\cdots i_{m}}$,$1\leq
i_{1}\leq N_{1}, 1\leq i_{2}\leq N_{2},\ldots, 1\leq i_{m}\leq
N_{m}$ as a homogeneous coordinate-function on
$\mathbf{CP}^{N_{1}N_{2}\cdots N_{m}-1}$. Now, let us consider the
composite quantum system
$\mathcal{Q}^{p}_{m}(N_{1},N_{2},\ldots,N_{m})$ and let
\begin{equation}
\mathcal{A}=\left(\alpha_{i_{1},i_{2},\ldots,i_{m}}\right)_{1\leq
i_{j}\leq N_{j}},
\end{equation}
for all $j=1,2,\ldots,m$. $\mathcal{A}$ can be realized as the
following set $\{(i_{1},i_{2},\ldots,i_{m}):1\leq i_{j}\leq
N_{j},\forall~j\}$, in which each point $(i_{1},i_{2},\ldots,i_{m})$
is assigned the value $\alpha_{i_{1},i_{2},\ldots,i_{m}}$. This
realization of $\mathcal{A}$ is called an $m$-dimensional box-shape
matrix of size $N_{1}\times N_{2}\times\cdots\times N_{m}$, where we
associate to each such matrix a sub-ring
$\mathrm{S}_{\mathcal{A}}=\mathbf{C}[\mathcal{A}]\subset\mathrm{S}$,
where $\mathrm{S}$ is a commutative ring over the complex number
field. For each $j=1,2,\ldots,m$, a two-by-two minor about the
$j$-th coordinate of $\mathcal{A}$ is given by
\begin{eqnarray}\label{segreply1}
\mathcal{P}_{k_{1},l_{1};k_{2},l_{2};\ldots;k_{m},l_{m}}&=&
\alpha_{k_{1},k_{2},\ldots,k_{m}}\alpha_{l_{1},l_{2},\ldots,l_{m}}
\\\nonumber&&-
\alpha_{k_{1},k_{2},\ldots,k_{j-1},l_{j},k_{j+1},\ldots,k_{m}}\alpha_{l_{1},l_{2},
\ldots,l_{j-1},k_{j},l_{j+1},\ldots,l_{m}}\in
\mathrm{S}_{\mathcal{A}}.
\end{eqnarray}
Then the ideal $\mathcal{I}^{m}_{\mathcal{A}}$ of
$\mathrm{S}_{\mathcal{A}}$ is generated by
$\mathcal{P}_{k_{1},l_{1};k_{2},l_{2};\ldots;k_{m},l_{m}}$  and
describes the separable states in $\mathbf{CP}^{N_{1}N_{2}\cdots
N_{m}-1}$ \cite{Grone}. The image of the Segre embedding
$\mathrm{Im}(\mathcal{S}_{N_{1},N_{2},\ldots,N_{m}})$, which again
is an intersection of families of quadric hypersurfaces in
$\mathbf{CP}^{N_{1}N_{2}\cdots N_{m}-1}$, is given by
\begin{eqnarray}\label{eq: submeasure}
\mathrm{Im}(\mathcal{S}_{N_{1},N_{2},\ldots,N_{m}})&=&\bigcap_{\forall
j}\mathcal{V}\left(\mathcal{C}_{k_{1},l_{1};k_{2},l_{2};\ldots;k_{m},l_{m}}\right).
\end{eqnarray}
Moreover, we can define an entanglement measure for a pure
multipartite state as
\begin{eqnarray}\label{EntangSeg}
\mathcal{E}(\mathcal{Q}^{p}_{m}(N_{1},\ldots,N_{m}))&=&\left(\mathcal{N}\sum_{\forall
j}\left|\mathcal{P}_{k_{1},l_{1};k_{2},l_{2};\ldots;k_{m},l_{m}}\right|^{2}\right)^{\frac{1}{2}}
\\\nonumber
&=&(\mathcal{N}\sum_{\forall k_{j},l_{j},
j=1,2,\ldots,m}|\alpha_{k_{1},k_{2},\ldots,k_{m}}\alpha_{l_{1},l_{2},\ldots,l_{m}}
\\\nonumber&&-
\alpha_{k_{1},k_{2},\ldots,k_{j-1},l_{j},k_{j+1},\ldots,k_{m}}\alpha_{l_{1},l_{2},
\ldots,l_{j-1},k_{j},l_{j+1},\ldots,l_{m}}|^{2})^{\frac{1}{2}},
\end{eqnarray}
where $\mathcal{N}$ is an arbitrary normalization constant and
$j=1,2,\ldots,m$. This measure coincides with the generalized
concurrence for a general bipartite and three-partite state, but for
reasons that we have explained in \cite{Hosh5}, it fails to quantify
the entanglement for $m\geq 4$, whereas it still provides the
condition of full separability. However, it is still possible to
define an entanglement measure for general multipartite states if we
modify the equation (\ref{EntangSeg}) in such away that it contains
all possible permutations of indices. To do so, we propose a measure
of entanglement for general pure multipartite states as
\begin{eqnarray}\label{EntangSeg2}\nonumber
\mathcal{F}(\mathcal{Q}^{p}_{m}(N_{1},\ldots,N_{m}))
&=&(\mathcal{N}\sum_{\forall \text{Perm} (\sigma)}\sum_{\forall
k_{j},l_{j},
j=1,2,\ldots,m}|\alpha_{k_{1},k_{2},\ldots,k_{m}}\alpha_{l_{1},l_{2},\ldots,l_{m}}
\\&&-
\alpha_{\sigma(k_{1}),\sigma(k_{2}),\ldots,\sigma(k_{m})}\alpha_{\sigma(l_{1}),\sigma(l_{2}),
\ldots,\sigma(l_{m})}|^{2})^{\frac{1}{2}},
\end{eqnarray}
where $\text{Perm}(\sigma)$ denotes all possible permutations of
indices  $k_{1},k_{2},\ldots,k_{m}$ by $l_{1},l_{2}, \ldots,l_{m}$,
e.g., the first set of permutations is give by equation
(\ref{segreply1}) defining the Segre variety. By construction this
measure of entanglement vanishes on product states and it is also
invariant under all possible permutations of indices. Note that the
first set of permutations defines the Segre variety, but there are
also additional products of the complex projective spaces as
subspaces of $\mathbf{CP}^{N_{1}N_{2}\cdots N_{m}-1}$ which are
defined by other sets of permutations of indices in equation
(\ref{EntangSeg2}).


\section{Some examples: three-partite and four-partite states}
 In this section we will apply this measure of entanglement to
 three-partite and four-partite states and give explicit expression
 for the measure of entanglement for these states. We start by the
 simplest multipartite states, namely three-partite states.
 Following the recipe in the general expression for multipartite
 states, we can write the measure of entanglement for such states as
\begin{eqnarray}\label{threepartie}
\mathcal{F}(\mathcal{Q}^{p}_{3}(N_{1},N_{2},N_{3}))&=&(\mathcal{N}\sum_{\forall
\text{Perm} (\sigma)}\sum_{\forall k_{j},l_{j},
j=1,2,3}|\alpha_{k_{1},k_{2},k_{3}}\alpha_{l_{1},l_{2},l_{3}}\\\nonumber&&-
\alpha_{\sigma(k_{1}),\sigma(k_{2}),\sigma(k_{3})}\alpha_{\sigma(l_{1}),\sigma(l_{2}),
\sigma(l_{3})}|^{2})^{\frac{1}{2}}\\\nonumber &=&
(\mathcal{N}\sum^{3}_{p_{1}=1}\sum_{\forall
k_{j},l_{j}}|\alpha_{k_{1},k_{2},k_{3}}\alpha_{l_{1},l_{2},l_{3}}-
\alpha_{k_{1},l_{p_{1}},k_{3}}\alpha_{l_{1},k_{p_{1}},l_{3}}|^{2})^{\frac{1}{2}}\\\nonumber
&=&(\mathcal{N}\sum_{k_{1},l_{1};k_{2},l_{2};k_{3},l_{3}}
(\left|\alpha_{k_{1},k_{2},k_{3}}\alpha_{l_{1},l_{2},l_{3}}-
\alpha_{k_{1},k_{2},l_{3}}\alpha_{l_{1},l_{2},k_{3}}\right|^{2}\\\nonumber&&
+\left|\alpha_{k_{1},k_{2},k_{3}}\alpha_{l_{1},l_{2},l_{3}}-
\alpha_{k_{1},l_{2},k_{3}}\alpha_{l_{1},k_{2},l_{3}}\right|^{2})\\\nonumber&&+
\left|\alpha_{k_{1},k_{2},k_{3}}\alpha_{l_{1},l_{2},l_{3}}-
\alpha_{l_{1},k_{2},k_{3}}\alpha_{k_{1},l_{2},l_{3}}\right|^{2})^{\frac{1}{2}}.
\end{eqnarray}
This measure of entanglement for three-partite states
(\ref{threepartie}) coincides with generalized concurrence
\cite{Albeverio}. Moreover, this measure of entanglement measure is
equivalent but not equal to our entanglement tensor based on joint
POVMs on phase space \cite{Hosh4}. Next, we will discuss the measure
of entanglement for four-partite states. In this case, we have more
then one set of permutations and as we have explained before this is
the reason why the measure of entanglement that is directly based on
the polynomial that define the Segre variety fails to quantify the
entanglement of four-partite states. Now, a measure of entanglement
based on the modified Segre variety  for four partite states is
given by
\begin{eqnarray}\label{fourpartie}\nonumber
&&\mathcal{F}(\mathcal{Q}^{p}_{4}(N_{1},N_{2},N_{3},N_{4}))=(\mathcal{N}\sum_{\forall
\text{Perm} (\sigma)}\sum_{\forall k_{j},l_{j},
j=1,2,3,4}|\alpha_{k_{1},k_{2},k_{3},k_{4}}\alpha_{l_{1},l_{2},l_{3},l_{4}}\\\nonumber&&-
\alpha_{\sigma(k_{1}),\sigma(k_{2}),\sigma(k_{3}),\sigma(k_{4})}\alpha_{\sigma(l_{1}),\sigma(l_{2}),
\sigma(l_{3}),\sigma(l_{4})}|^{2})^{\frac{1}{2}}\\\nonumber &&
(\mathcal{N}[\sum^{4}_{p_{1}=1}\sum_{\forall
k_{j},l_{j}}|\alpha_{k_{1},k_{2},k_{3},k_{4}}\alpha_{l_{1},l_{2},l_{3},l_{4}}-
\alpha_{k_{1},k_{2},l_{p_{1}},k_{4}}\alpha_{l_{1},l_{2},k_{p_{1}},l_{4}}|^{2}
\\ &&+
\sum_{p_{1}<p_{2}}\sum_{\forall
k_{j},l_{j}}|\alpha_{k_{1},k_{2},k_{3},k_{4}}\alpha_{l_{1},l_{2},l_{3},l_{4}}-
\alpha_{k_{1},l_{p_{1}},l_{p_{2}},k_{4}}\alpha_{l_{1},k_{p_{1}},k_{p_{2}},l_{4}}|^{2}]
)^{\frac{1}{2}}.
\end{eqnarray}
The first sum in equation (\ref{fourpartie}) defines the Segre
variety, and the second sum  gives  another product of the complex
projective spaces which is still a quadratic subspace of
$\mathbf{CP}^{N_{1}N_{2}N_{3} N_{4}-1}$. For example, for four-qubit
states the first set of permutations represented by the first sum
gives 112 quadric terms and the second set of permutations, which is
represented by the second sum, gives 36 quadric terms. Thus a
measure of entanglement for four-qubit states contains 148 terms.
\section{Conclusion}

In this paper, we  have constructed a geometric well-motivated
measure of entanglement for general pure multipartite  states,
 based on an extension of the Segre variety. This measure of
 entanglement works for any  pure state and vanishes on
 multipartite
 product states. We have also given explicit expressions for our
 entanglement measure for general pure three-partite and four-partite
 states. Moreover, we have compared this measure of entanglement
 with the generalized concurrence.

\begin{flushleft}
\textbf{Acknowledgments:} The  author gratefully acknowledges the
financial support of the Japan Society for the Promotion of Science
(JSPS).
\end{flushleft}


\end{document}